\documentclass[aps,prb,twocolumn,groupedaddress,showpacs,showkeys]{revtex4-1}

\usepackage{graphicx}
\usepackage{multirow}

\bibstyle{apsrev}

\begin{document}

\title{Anisotropic strain in SmSe and SmTe: implications for electronic transport}

\author{Marcelo A. Kuroda$^{1,2}$}
\email{mkuroda@auburn.edu}
\author{Zhengping Jiang$^3$} \author{ Michael Povolotskyi$^3$} \author{  Gerhard Klimeck$^3$} \author{ Dennis M.~Newns$^1$ } \author{  Glenn J.~Martyna$^1$}
\affiliation{$^1$ IBM T.~J.~Watson Research Center, Yorktown Heights, NY 10598, USA\\$^2$ Department of Physics, Auburn University, Auburn, AL 36849, USA\\ $^3$ Network for Computational Nanoelectronics, School of Electrical and Computer Engineering, Purdue University, West Lafayette, IN 47906, USA}
\begin{abstract}
Mixed valence rare-earth samarium compounds SmX (X=Se,Te) have been recently proposed as candidate materials for use in high-speed, low-power digital switches driven by stress induced changes of resistivity.  At room temperature these materials exhibit a pressure driven insulator-to-metal transition with resistivity decreasing by up to 7 orders of magnitude over a small pressure range. Thus, the application of only a few GPa's to the piezoresistor (SmX) allows the switching device to perform complex logic.  Here we study from first principles the electronic properties of these compounds under uniaxial strain and discuss the consequences on carrier transport.  The changes in the band structure show that the piezoresistive response is mostly governed by the reduction of band gap with strain.  Furthermore, it becomes optimal when the Fermi level is pinned near the localized valence band.  The piezoresistive effect under uniaxial strain which must be taken into account in thin films and other systems with reduced dimensionality is also quantified.  Under uniaxial strain we find that the piezoresistive response can be substantially larger than in the isotropic case.  Analysis  of complex band structure of SmSe yields a tunneling length of the order of 1 nm.  The results suggest that the conduction mechanism governing the piezoresistive effect in bulk, i.e.~thermal promotion of electrons, should still be dominant in few-nanometer-thick films.
\end{abstract}

\date{\today}


\maketitle

\section{Introduction}

There is currently great interest in novel electronic switches for use in post-CMOS logic \cite{theis2010}. Recently, a new high performance computer switch has  been proposed promising high speed at low power \cite{newns2012b}. The switch operates through the use of a nanoscale piezoelectric actuator applying uniaxial stress to a piezoresistive element.  The piezoresistive material undergoes an insulator to metal transition thereby turning the device on.    The device can be described as a type of nano-electro-mechanical system (NEMS) device wherein all the parts stay in continuous mechanical contact.

In order to form an effective device, the piezoelectric and the piezoresistor must exhibit high performance. Here we concentrate on the piezoresistor which must have high sensitivity to applied strain (large resistance drop, $\sim10^4$, with little applied stress), fast response to the applied stress and high endurance to cycling.  These requirements define the engineering envelope for effective piezoresistive materials in logic devices \cite{barlian2009}.  For example, assemblies of coated nanoparticles \cite{athanassiou2006} and polymeric fiber arrays \cite{pang2012} exhibit large piezoresistive response.  However, such systems will neither scale nor withstand cycling and large current densities like truly solid state systems \cite{imada1998} such as vanadium oxides (Cr-doped V$_2$O$_3$ \cite{mcwhan1969} and VO$_2$ \cite{marezio1972}) or mixed valence rare-earth monochalcogenide compounds (e.g.~SmSe, SmTe, TmTe)\cite{jayaraman1970,jayaraman1970b,batlogg1976, matsumura1996}.  The piezoresistive effect in VO$_2$  systems occur via a structural transition and may be hysteretic  \cite{cavalleri2004} and hence slow or irreversible. In Cr-doped V$_2$O$_3$, experimental work has shown that transitions may be suppressed in thin films \cite{luo2004}.  In contrast samarium monochalcogenide (SmX with X = Se, Te) compounds can switch easily and reversibly as the insulator-to-metal transition is continuous, isostructural and importantly for nanoscale devices has been recently observed in thin films \cite{copel2013}.

The continuous insulator-to-metal transition in bulk crystalline rare-earth monochalcogenide materials has been experimentally studied beginning with the pioneering diamond anvil cell work of Jayaraman \textit{et al} in the 1970's \cite{jayaraman1970, batlogg1976, matsumura1996}.  The piezoresistive effect in these materials shows several orders of magnitude change in resistance with a constant piezoresistive gauge (the logarithmic derivative of resistance with respect to pressure).  Previous theoretical calculations using density functional theory (DFT)  \cite{farberovich1980, antonov2002, svane2004, gupta2009,petit2014} focused on the different phases of these materials.  These studies (both experimental and theoretical) have been limited to the analysis of effects of hydrostatic (i.e.~isotropic) strain.  However, the quantitative analysis of the piezoresistive response in SmX thin films \cite{copel2013} and their potential for nanoscale devices that rely on anisotropic strain demand further study of the effects of strain/stress on electronic properties and ultimately on transport.  Moreover, as systems sizes are reduced, it is important to determine when tunneling becomes relevant and thus modifies the bulk piezoresistive response.

The present work uses first principles calculations to quantify the changes in the electronic band structure in SmSe and SmTe under uniaxial strain and to develop a basic model which contains the essential physical describing piezoresistivity in the diffusive regime.  Piezoresistive response is in particular studied for different configurations of strain along the [001], [011] and [111] directions and compared to the isotropic case providing a more detailed framework to the recent experimental demonstration in SmSe thin films \cite{copel2013}.   Considerations for systems with ultra short channel length where the piezoresistive effect can be reduced by tunneling are also discussed by analyzing the complex band structure (CBS) via an empirical tight-binding model.  By providing a more complete picture of the strain effects in SmX systems (beyond the scope of previous experimental and theoretical works) this work offers important insights on the piezoresistive effect in these compounds.  In addition, the study is designed to identify different factors which at the nanoscale are of importance to proposed applications in low-power switching \cite{newns2012a,newns2012b}. 

The manuscript is organized as follows. We first present the pressure dependent properties that govern electronic transport in the diffusive regime under isotropic strain (Section~\ref{sec2}).  We next provide a comparison of the changes of those properties between isotropic and uniaxial strain (Section~\ref{sec3}). We analyze the complex band structure in different directions to estimate the length scale at which tunneling becomes important (Section~\ref{sec4}) before presenting our conclusions (Section~\ref{sec5}).

\section{\label{sec2} Diffusive transport regime}

The electronic properties of SmSe and SmTe are calculated using DFT under the generalized gradient approximation \cite{langreth1980,pbe96}.  In these compounds, the splitting of the Sm-4f band originates from both on-site (Hubbard-like) and spin-orbit interactions.  Calculations presented here include both contributions.  Electronic densities are obtained employing the full-potential linearized augmented plane-wave (FP-LAPW) method \cite{elkcode} with the Hubbard term computed in the fully-localized limit.  The value of on-site interaction $U = 6$ eV has been selected such that it captures the optical spectra of SmSe \cite{batlogg1976, uchoice}.  The calculated values for the elastic constants for SmSe and SmTe show good agreement with the available experimental data as indicated in Table \ref{tab:tab1}.

\begin{figure}[h]
\centering
\includegraphics[width= 3.in,clip=true]{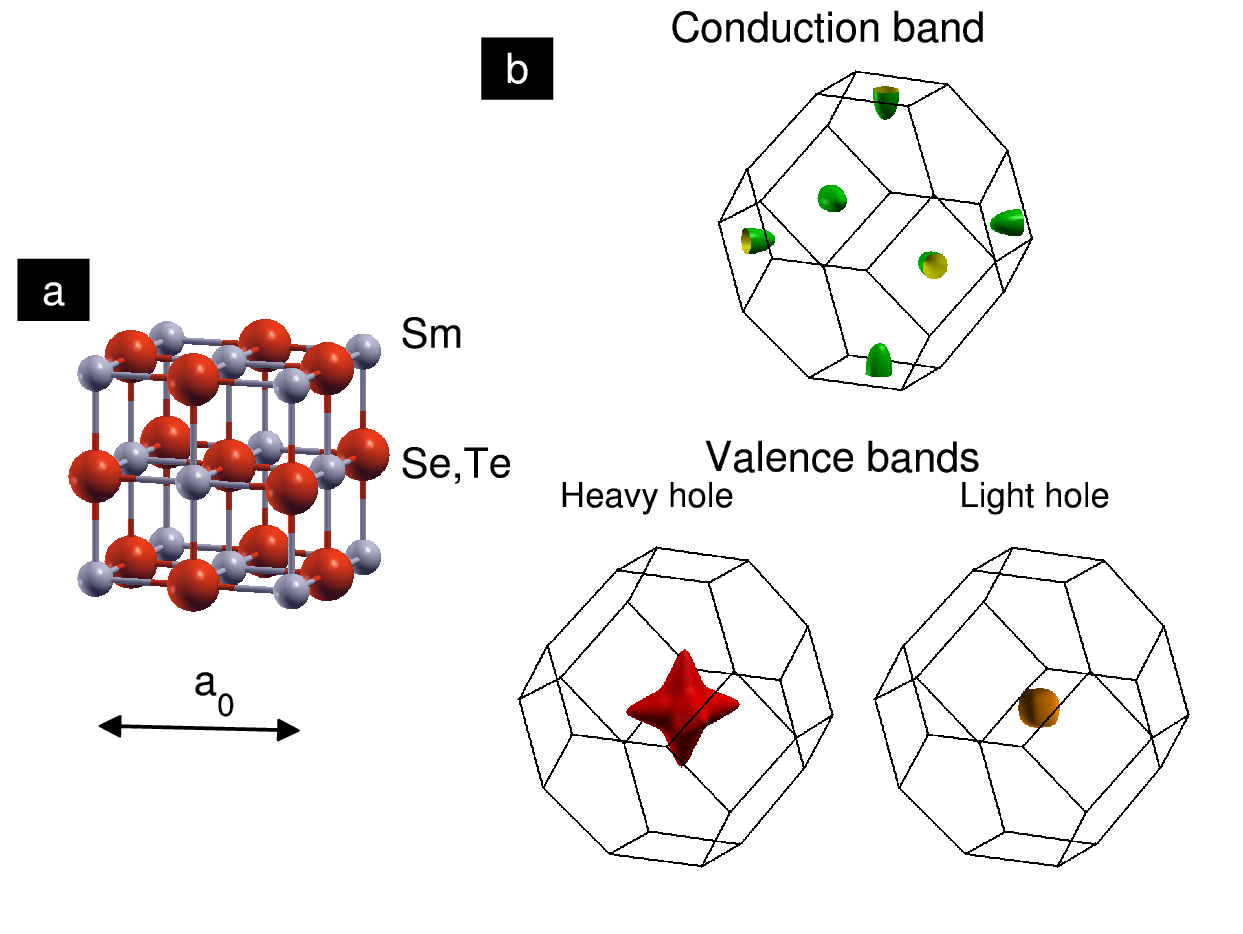}
\includegraphics[width= 3.25in]{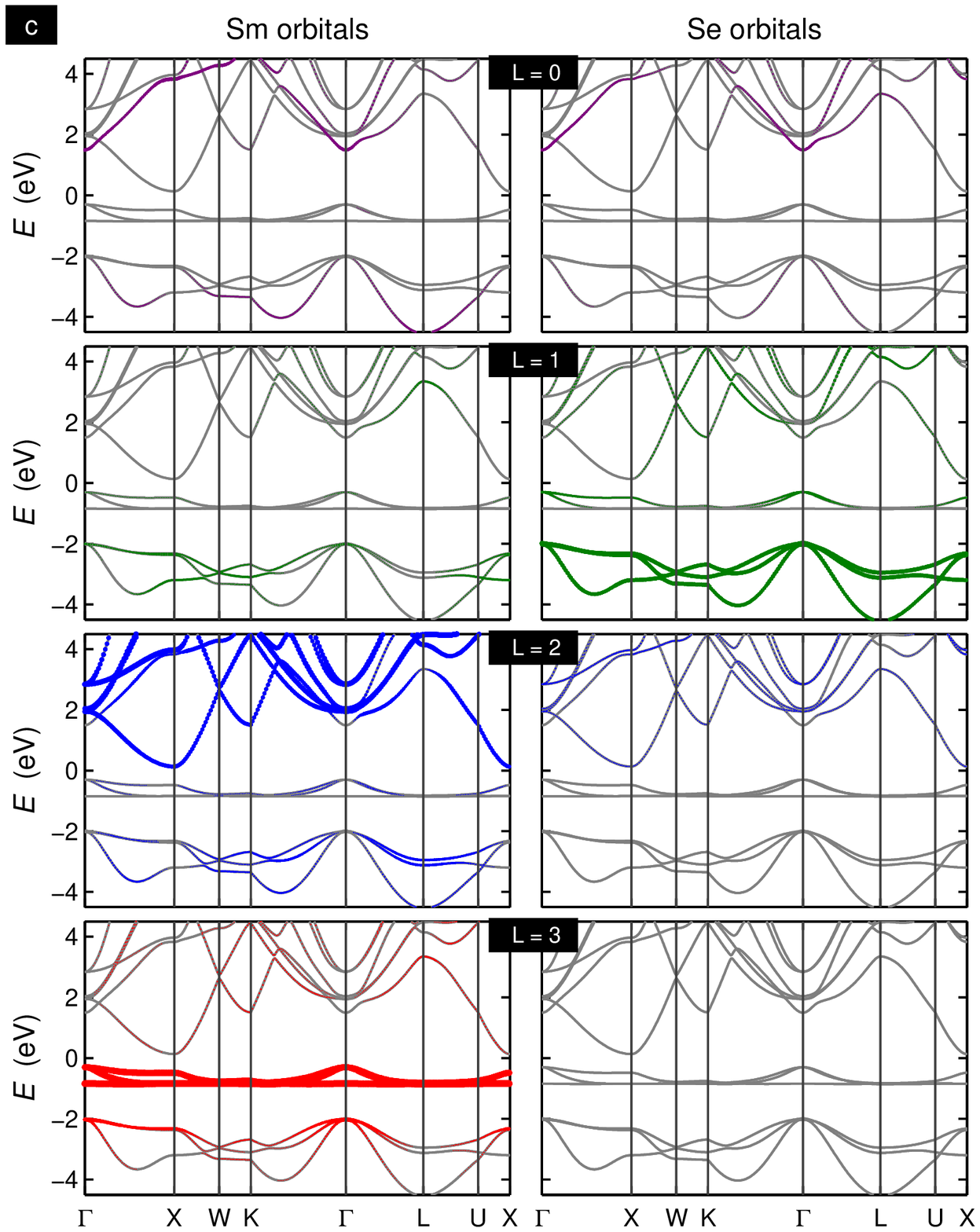}
\caption{(Color online) Crystal and band structure of rock-salt SmSe (a) Crystal structure.  (b) Energy surfaces of conduction (top) and valence bands (bottom) at 0.25 eV above and below the band edge, respectively.  (c) Band structure of SmSe near the Fermi level ($E_F = 0$).  Color indicates projections of Bloch states onto localized atomic orbitals of Sm (left) and Se (right) with different angular momentum (from top to bottom): $L$ = 0 (purple), 1 (green), 2 (red) and 3 (blue).}\label{fig:fig1}
\end{figure}

\begin{figure}[htbp]
\centering
\includegraphics[width= 3.25in]{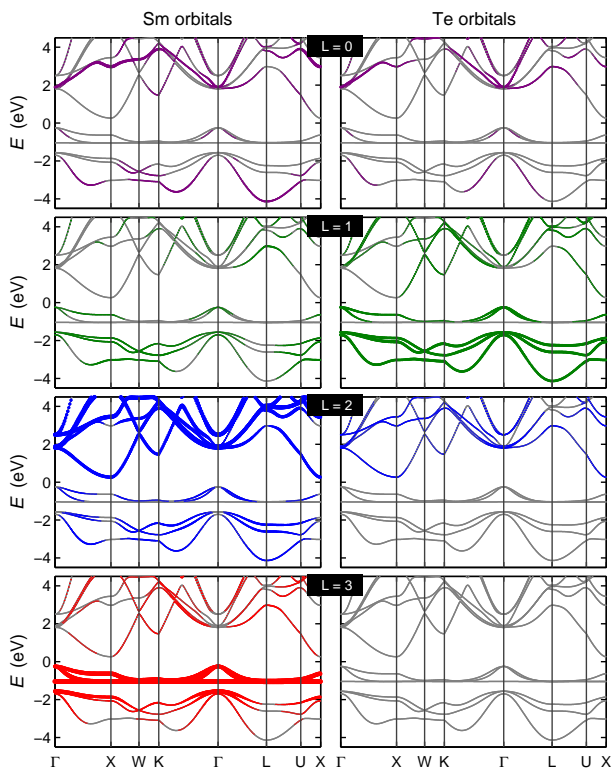}
\caption{(Color online) Band structure of rock-salt SmTe near the Fermi level ($E_F = 0$).  Color indicates projections of Bloch states onto localized atomic orbitals of Sm (left) and Te (right) with different angular momentum (color code as Fig.~\ref{fig:fig1}).}\label{fig:smte_bs}
\end{figure}

Mixed valence rare-earth compounds SmSe and SmTe have rock-salt crystal structure (Fm$\bar{3}$m), as depicted in Fig.~\ref{fig:fig1}a.  Band structure shows them to be indirect band gap semiconductors \cite{antonov2002, svane2004, gupta2009} with $E_g \sim 0.5$ eV.  The valence bands maximum is located at the $\Gamma$-point and their energy surfaces have octahedral symmetry (as illustrated for the case of SmSe in Fig.~\ref{fig:fig1}b).  In both systems the valence band is formed by the highly localized Sm-4f states ($L=3$) while the chalcogenide p-bands ($L=1$) reside below the localized Sm-4f band, as shown in Fig.~\ref{fig:fig1}c and \ref{fig:smte_bs}.  The conduction band edge, located at the X symmetry point, is mostly formed by the delocalized (mobile) Sm 5d states ($L=2$).  The indirect band gap occurs between the conduction band X point and the valence band $\Gamma$ point. The direct (optical) band gap $E_{op}$ is at the X point.  The energy surfaces around the X point in the conduction band are well described by ellipsoids (Fig.~\ref{fig:fig1}b) and, similarly to silicon, the transverse effective mass is smaller than the longitudinal one ($m^l_e/m^t_e \approx 5.5 $).

\begin{table}[h]
\caption{\label{tab:tab1}Physical properties for the cubic phase (Fm$\bar{3}$m) of SmSe and SmTe as obtained from first principles calculations: lattice constant ($a_0$), band gap ($E_g$), bulk modulus ($B_0$), Poisson ratio ($\nu$), elastic constants ($c_{11}$, $c_{12}$ and $c_{44}$) and band gap change with pressure ($dE_g\!/\!dp$). }
\begin{center}
\begin{tabular}{c c c c c}
& \multicolumn{2}{ c }{SmSe} &\multicolumn{2}{ c }{SmTe} \\ \cline{1-5}
Property & Theory & Expt. & Theory  & Expt.\\ \cline{1-5}
$a_0$ (\AA) &  6.18  & 6.20 \cite{jayaraman1970} & 6.578 & 6.595 \cite{chatterjee1972}\\
$E_g$ (eV) & 0.5  & 0.5 \cite{jayaraman1970}& 0.54  & 0.63\cite{jayaraman1970} \\
$B_0$ (GPa) &  38.3 & $40\pm5$  \cite{jayaraman1974}& 34.0& $40\pm5$\cite{chatterjee1972}\\
$\nu$ &  0.18 & -- & 0.18 & --\\
$c_{11}$ (GPa) & 93.7 & -- & 87.4&--\\
$c_{12}$ (GPa) & 10.9 & --& 7.3&-- \\
$c_{44}$ (GPa) &  24.2 &-- & 17.2&--\\
\multirow{1}{*}{$\begin{array}{c}
 dE_g\!/\!dp\\ \mbox{(eV/GPa)}\\
\end{array}$ } & -0.13 & -$0.11\!\pm\!0.01$\cite{kirk1972} & -0.13 & -$0.12\!\pm\!0.01$\cite{jayaraman1970}\\
\end{tabular}
\end{center}
\end{table}

Experimental work on these mixed-valence compounds suggests that conduction in these systems is due to electron carriers \cite{batlogg1976,matsumura1996}.  More recently further experimental work using X-ray-absorption spectroscopy found the thermal promotion of 4f electrons to conducting bands occur at low pressure \cite{jarrige2013} before hybridization  governs the valence transition.  Contributions from holes are omitted in spite of the finite bandwidth of the 4f valence band stemming from the mixing between Sm-4f and chalcogenide 5p states (Figs.~\ref{fig:fig1}c and 2).  We therefore attribute the absence of hole conduction to strong scattering in the Sm 4f band or to an enhancement of the band mass due to strong correlations.  The effect of mass enhancement can be inferred from data at low temperatures \cite{maple1971,bader1973,varma1976} where the 4f band width is considerably narrower ($\lesssim 50$ meV) than the first principles calculations, suggesting strong localization of the 4f states in SmX materials (S, Se, Te).

The piezoresistive effect, which is also observed in semiconductors such as silicon or germanium \cite{smith1954},  is particularly large in samarium chalcogenide compounds.  The large deformation potential \cite{bardeen1950} rapidly reduces the band gap between the Sm-4f and Sm-5d bands under compressive strain.  Consequently electrons are thermally promoted from the localized Sm-4f valence bands to the mobile Sm-5d conduction bands and a continuous (isostructural \cite{svane2004, gupta2009,jarrige2013}) insulator-to-metal transition arises.

\begin{figure}[htpb]
\centering
\includegraphics[width= 3.25in]{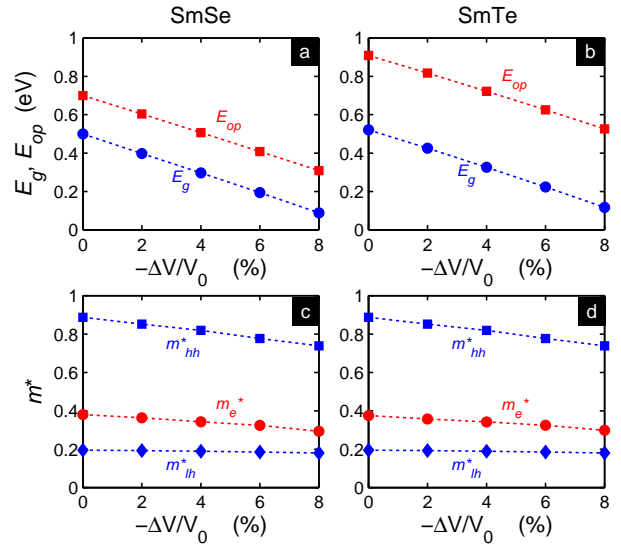}
\caption{(Color online) Changes in the electronic properties with hydrostatic strain.  Indirect (circles) and optical (squares) band gaps as a function of the (isotropic) volume change: (a) SmSe; (b) SmTe.   Effective masses of bands near the Fermi level: light holes (diamonds), heavy holes (squares), and electrons (circles) as a function of the (isotropic) volume change: (c) SmSe; (d) SmTe.}\label{fig:pressure_dep}
\end{figure}

A quantitative description of the piezoresistive effect requires the characterization of changes in the band structure (band gap and effective masses) under strain.  Figures \ref{fig:pressure_dep}a and \ref{fig:pressure_dep}b show the indirect (between $\Gamma$ and X) and direct (at symmetry point X) electronic band gaps as a function of change in volume under isotropic compressive strain in SmSe and SmTe, respectively.  The indirect band gap $E_g$ as obtained from first principles is similar for both cases while the optical gap $E_{op}$ is larger in SmTe than in SmSe. However, they both change at the same rate with strain as observed experimentally \cite{jayaraman1970}.  The carrier effective masses \cite{szebook} [$m^*=(m^*_{1}m^*_{2}m^*_{3})^{1/3}$] of each band  near the Fermi level are also computed: the light hole (lh) and heavy hole (hh) at the $\Gamma$ point and the electron at the X point.  The results show that values are similar for both systems and small reduction with compressive strain (Fig.~\ref{fig:pressure_dep}c and \ref{fig:pressure_dep}d).

The piezoresistive gauge is defined as \cite{mason1957}:
\begin{equation}
\pi_x = \frac{d}{dx}\log\sigma \label{eq:pr_gauge},
\end{equation}
where $\sigma$ is the conductivity and $x$ is the field (e.g. strain or pressure).  In this work we focus mostly on volumetric strain (i.e.~$x=\Delta V/V_0$).  Under isotropic strain, pressure and volumetric strain piezoresistive gauges are related by:  
\begin{equation}
\pi_p = \frac{\pi_x}{B_0},
\end{equation}
where $B_0$ is the bulk modulus.  In the diffusive regime, the conductivity can be approximated as modulated by the pressure-dependent charge carrier population in the 5d-conduction band:
\begin{equation}
\sigma = e n_e \mu_e
\end{equation}
where $n_e$ and $\mu_e$ are the electron density and mobility, respectively. The electron density $n_e$ is expressed as:
\begin{equation}
n_e(x) = N_c \exp\left(-E_a/k_BT\right), \label{eq:electron_density}
\end{equation}
where $N_c = 6 [(m^*_e k_BT)/(2\pi\hbar^2)]^{1/3}$ and $E_a = E_c-E_F$ are the density of states and activation energy (i.e.~position of the conduction band relative to the Fermi level), respectively.  In Eq.~\ref{eq:electron_density} we left implicit the dependence of $N_c$ and $E_a$ on strain.

Neglecting changes with strain in scattering mechanisms the piezoresistive gauge (Eq.~\ref{eq:pr_gauge}) can then be approximated by:
\begin{equation}
\pi_x = -\frac{1}{k_BT}\frac{d}{dx}E_a+\frac{3}{2}\frac{d}{dx} \left(\log m^*_e\right). \label{eq:pr_gauge2}
\end{equation}
From the results obtained from first principles calculations (Fig.~\ref{fig:pressure_dep}) we find that the second term on the right hand side of Eq.~\ref{eq:pr_gauge2} is more than an order of magnitude smaller than the first one.  Therefore the piezoresistive gauge in the Sm monochalcogenide systems is largely governed by the reduction of the activation energy with strain at room temperature or below.  Under such conditions, the piezoresistive gauge (Eq.~\ref{eq:pr_gauge2}) can be estimated using:
\begin{equation}
\pi_x = -\frac{1}{k_BT}\frac{dE_a}{dx}. \label{eq:pr_gauge_appr}
\end{equation}

We evaluate the piezoresistive response of SmSe and SmTe in two different scenarios using Eq.~\ref{eq:electron_density}: (i) single crystal under isotropic strain, and (ii)  pinned $E_F$.  In the \emph{ideal} single crystal case under isotropic strain, the Fermi level position can be obtained analytically using the effective masses for each band \cite{uniaxial}.
Using the linear dependence on strain obtained from first principles calculations, we find that
the piezoresistive gauge can be approximated as
\begin{equation}
\pi_p^{bulk} \sim -\frac{ 1}{2k_BT} \frac{dE_g}{dp} \approx 2.6 \mbox{ GPa}^{-1}
\end{equation}
for these material systems at room temperature.

In reality the presence of defects or impurities levels can place Fermi level away from the middle of the band gap as in ideal crystalline systems.  In thin films the Fermi level can also be pinned by the metal contacts when distance between electrodes becomes comparable to the screening length allowing to tailor the piezoresistive gauge in thin films (at larger separations, Fermi-level pinning it is likely controlled strongly by the metal-induced-gap states \cite{tersoff1984}).   For cases where the Fermi-level pinning is unknown, an upper bound value for piezoresistive gauge can be determined:
\begin{equation}
\pi_s \lesssim \frac{1}{k_BT}\left|\frac{dE_g}{ds}\right|, \label{eq:lower_bound}
\end{equation}	
corresponding to the case where the Fermi level sits at the edge of the valence band.  Under isotropic strain, the upper bound at room temperature becomes:
\begin{equation}
\pi_p  \approx 5.1 / \mbox{GPa},
\end{equation}
which is very close to the values of 5.3 and 4.9 /GPa for SmSe and SmTe respectively reported by Jayaraman \textit{et al} \cite{jayaraman1970,notesmte}.  We henceforth conclude that those results under hydrostatic strain are likely near the best possible case scenario within the physics accounted for in our transport model. Because the linear dependence of band gap with strain is a good approximation for large strain (up to $|\Delta V/V_0| \sim 8$ \%) the piezoresistive gauge remains nearly constant over a large stress/strain range, as observed experimentally.

\section{\label{sec3} Uniaxial strain} 
\begin{figure}[htpb]
\centering
\includegraphics[width= 3.4in]{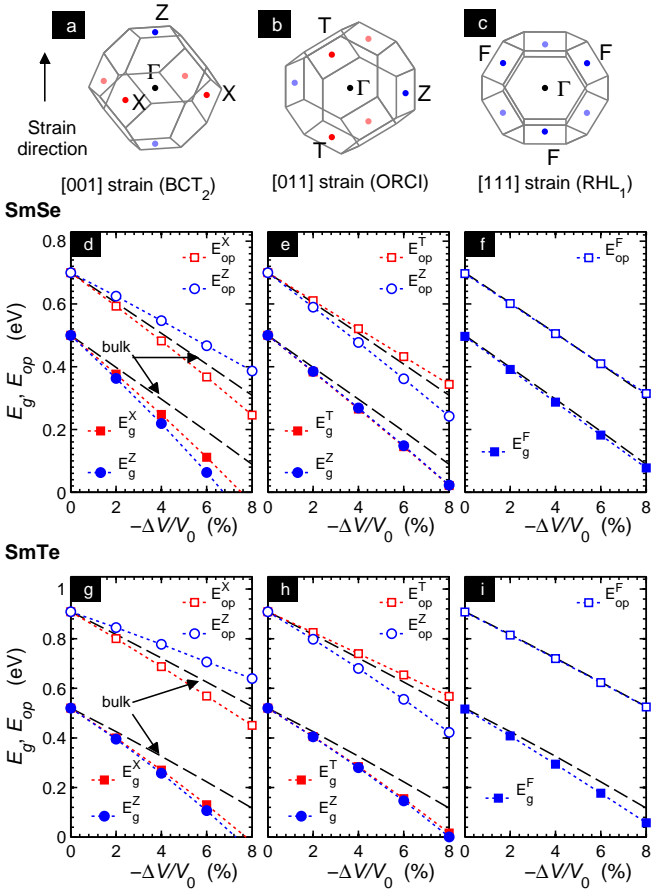}
\caption{(Color online) Band gap reduction under uniaxial strain in SmX.  (Top row) 3D Brillouin zone of the rock-salt crystal structure Fm$\bar{3}$m turns into one with lower symmetries under uniaxial strain: body centered tetragonal BCT$_2$ ([001]), body-centered orthorhombic ORCI ([011]) and rhombohedral RHL ([111]).  In the first two cases, degeneracy at symmetry points where the conduction band edge resides is lifted. The center and bottom rows indicate change of the indirect (solid symbols) and optical (open symbols) band gap as a function of the volume change for SmSe and SmTe, respectively.  Each column corresponds to the strain case indicated in the top row. For comparison we also show the results for isotropic strain (dashed lines).  The top of the valence bands resides in all cases at the $\Gamma$ symmetry point; bottoms of the conduction band are located at the X symmetry point for the FCC, at X and Z for the BCT$_2$, at T and Z for the ORCI, and F for the RHL.}\label{fig:fig3_uniaxial_strain}
\end{figure}

In order to analyze the electrical response to stress in SmX thin films \cite{copel2013} and their potential application in high-performance NEMS such as the piezoelectronic transistor \cite{newns2012a}, the early work on hydrostatic strain must be significantly extended to consider the effects of uniaxial strain.  Thus we next quantify the changes in the band structure in SmSe and SmTe for uniaxial strain along the following three directions: [001], [011]  and [111], and compare them to the isotropic compression.  Note that uniaxial strain transforms the FCC structure (Fm$\bar{3}$m symmetry) into body centered tetragonal (BCT$_2$), body-centered orthorhombic (ORCI) and rhombohedral (RHL) structures for uniaxial strain along the [001], [011]  and [111] directions, respectively.   As a consequence of the reduced symmetry degeneracy is lifted and the electronic properties may change differently in different symmetry points (Fig.~\ref{fig:fig3_uniaxial_strain}).

In the diffusive regime the piezoresistive response can still be described using Eq.~\ref{eq:pr_gauge2}.  Therefore, differences in the piezoresistive gauge under uniaxial strain can be extracted from the strain dependence of band structure properties in that equation.  In particular the sensitivity of the band gap to strain which we have showed earlier to be the most prominent factor.

For the uniaxial strain cases considered here the one along the [001]-direction shows the largest change in indirect band gap ($E_g$), which is about 30\% larger than the isotropic case (Fig.~\ref{fig:fig3_uniaxial_strain}d and \ref{fig:fig3_uniaxial_strain}g).  Note that while pressure (the average of the stress tensor trace) is the same in these two cases, the enhancement emerges at the expense of larger axial component of stress due to the difference between $c_{11}$ and $c_{12}$ (Table~\ref{tab:tab1}).  The reduction in the indirect band gap for strain along the [011] direction is yet larger than that along the [111] direction.  In the [111] case, direct and indirect band gaps show similar values to those corresponding to hydrostatic strain.  For the cases of uniaxial strain along the [001] and [011] directions the band gaps in different symmetry points exhibit changes at different rates because of the strain-induced symmetry breaking.  This effect \cite{smith1954,kleinman1962} is known to not only change the band structure but also impact the mobility and has been exploited in strained Si/Ge interfaces \cite{vdwalle1986} and characterized using first principles calculations \cite{fischetti1996}.  However, relative changes in mobility will only be secondary (Eqs.~\ref{eq:pr_gauge} and \ref{eq:pr_gauge2}) compared to any change by the band gap.

\begin{figure}[htpb]
\centering
\includegraphics[width= 3.3in]{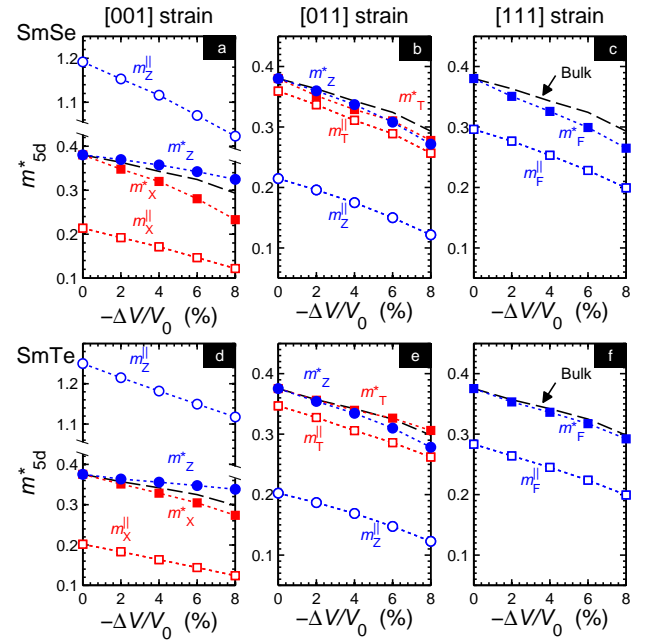}
\caption{(Color online) Effective mass dependence of conduction band under uniaxial strain in:  SmSe (top row) and SmTe (bottom row).  Uniaxial strain is applied along the [001] (left, not broken vertical axis), [011] (center) and [111] right. The effective mass along the strain direction (open symbols) can be significantly different than the effective mass (solid symbol) depending on the strain direction.  For comparison we also show the results for isotropic strain (dashed lines).}\label{fig:fig4_effmass}
\end{figure}

The dependence of the effective mass $m^*$ of the 5d bands (solid symbols in Fig.~\ref{fig:fig4_effmass}) shows minor differences with the hydrostatic case for both SmSe and SmTe.  Therefore it is expected that the piezoresistive gauge under uniaxial strain will still be dominated by the band gap reduction (first term in Eq.~\ref{eq:pr_gauge2}) and changes of band effective masses playing only a secondary role.  In contrast the band mass \cite{bandmass}  in the direction parallel to the strain direction (open symbols) can be considerably different to the effective mass (solid symbols).  For instance the band mass along the [001] direction ($m^{\parallel}$) can be considerably different to the effective mass ($m^*$) due to the large mass anisotropy.  This result has important implications in systems with short channel length as discussed later.

The first demonstration of the piezoresistive effect in thin films \cite{copel2013} achieved up to nearly half the bulk value in films ranging from 8 up to 50 nm.  This less than optimal performance can be attributed to several reasons such as the polycrystalline sample morphology with grain sizes tens of nanometers or the effect of metal electrodes mentioned above.  As film quality improves moving toward epitaxial samples a combination of uniaxial strain and proper choice of metal electrodes can be exploited to attain a larger piezoresistive gauge in NEMS applications.  Figure 4 suggests that the piezoresistive gauge could be enhanced as much as 30\% because of the stronger reduction in band gap with strain.

\section{\label{sec4} Tunneling}

As device sizes shrink, tunneling becomes relevant to conduction.  In such scenario crystal orientation exhibits considerably different band masses and thus impacts transport in devices with an ultra short channel.  The tunneling current $J$ can be estimated following Landauer formalism \cite{dicarlo1994}:
\begin{equation}
J = \frac{e}{(2\pi)^2 \hbar }\int\limits_{BZ_{2D}} d\mathbf{k} \int T(E,\mathbf{k})  \left[f_R(E)-f_L(E) \right] dE
\end{equation}
where $\mathbf{k}$ is the 2-dimensional wave vector in the Brillouin zone perpendicular to the transport direction, and $f_{R}(E)$ and $f_L(E)$ are the Fermi-Dirac distribution functions on the electrodes.  

As a first approximation we can estimate tunneling length in SmX from its CBS neglecting effects of the interface with the electrode.  The transmission probability is then taken to be:
\begin{equation}
T(E,\mathbf{k}) \propto \exp[- 2 \,\mbox{Im}(k_\parallel) t_{\mathrm{SmX}}]\label{eq:rate}
\end{equation}
where Im($k_\parallel$) is the imaginary part of the complex wave vector parallel to the strain direction and $t_{\mathrm{\small{SmX}}}$ is the thickness of the SmX films.  A more accurate estimate of electrical systems such system requires a complete atomistic description of the metal/SmX interface and is beyond the scope of this work.

\begin{figure}[htpb]
\centering
\includegraphics[width= 3.35in]{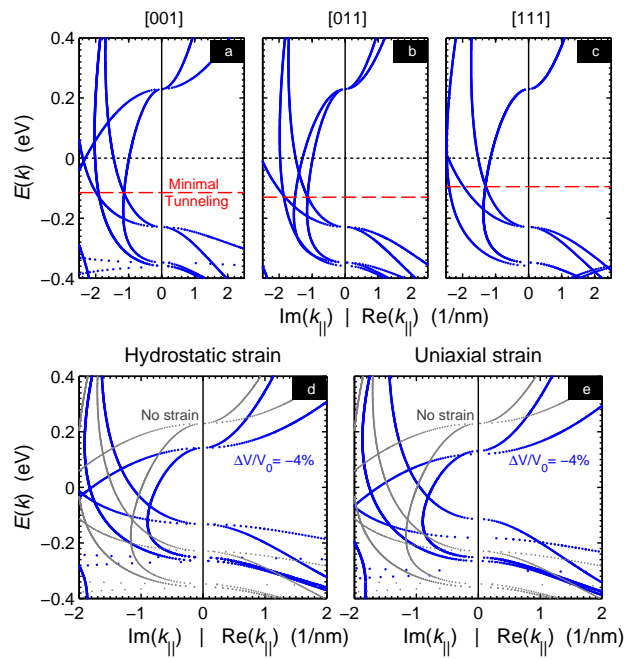}
\caption{(Color online) Complex band structure (CBS) of SmSe near the Fermi level at symmetry points of the 2D Brillouin zone. Real and imaginary band structures are plotted along the positive and negative wave vector axis, respectively.  (Bands appear to have direct band gaps due to band folding and superposition.)  We use as energy reference the middle of the indirect band gap (dotted line). Unstrained crystal for transport along different directions: (a) [001], (b) [011], and (c) [111].  The energy level at which tunneling is minimized (i.e. imaginary wave vectors are largest) resides below the middle of the gap (dashed line). Bottom row: CBS along the [001] direction under 4\% compressive strain: (d) hydrostatic and (e) uniaxial. The unstrained band structure is also plotted for reference (gray). }\label{fig:cbs}
\end{figure}

Figure~\ref{fig:cbs} shows the CBS of SmSe along the three directions previously considered [001], [011] and [111].  The results are obtained using an empirical tight-binding model accounting for second nearest neighbor coupling with full sets of ${spdf\!s}^*$ orbitals \cite{jiang2013}  employing NEMO5 \cite{steiger2011, fonseca2013}.  The CBS is plotted for the symmetry points where valence and conduction band edges are located.  

For a fixed energy $E$, a lower bound limit for tunneling (Eq.~\ref{eq:rate}) can be determined using the smallest imaginary part of the complex wave vector at that energy.  Overall tunneling in SmX thin films is minimized when the Fermi level position, which may be adjusted using the appropriate metal electrode, is located near the energy where the imaginary wave vectors are the largest.  From the CBS presented in Fig.~\ref{fig:cbs}, this occurs at an energy level above the valence band edge (as denoted by  dashed line).  The results indicate that Im($k_\parallel$) is of the order of 1~nm$^{-1}$, resulting in a nanometer tunneling length.  Thereby the governing conduction mechanism of the bulk (diffusive transport by thermally promoted electrons) is expected to remain dominant as long as the distance between electrodes is larger than the tunneling length.  Thus piezoresistive effect is expected to persist in crystalline films with thicknesses  $t_{\mathrm{\small{SmX}}}$ of a few nanometers \cite{copel2013}, with metal electrodes playing an important role.

Comparison between the unstrained CBS along different directions shows that the light electron bands may govern tunneling (shortest complex wave vector within the gap) and are approximately similar for all directions.  The [001] direction (top row of Fig.~\ref{fig:cbs}) shows very different imaginary bands due to the anisotropic mass of the 5d conduction band as previously shown in Fig.~\ref{fig:fig1}b.  The anisotropy is less significant along the [011] direction and not present for the [111] case. 

Study of the changes in the CBS upon the application of isotropic and uniaxial strain shows an overall reduction in the tunneling length.  As an example the case of  4\% hydrostatic and uniaxial compressive strain are shown in Fig.~\ref{fig:cbs}d and \ref{fig:cbs}e, respectively.  In all three strain cases, we observe a reduction in the complex wave vector in the band gap and the intersection between imaginary branches where the complex wave vector is the largest still occurs at energies below the middle of the band gap.  Similar results were found along the [011] and [111] directions (not plotted here).  The uniaxial strain produces a small splitting of the Sm-4f valence bands due to the lifting of degeneracy.  \\

\section{Conclusions\label{sec5} }

In summary we have studied the changes in the band structure of SmSe and SmTe under uniaxial strain using first principles calculations and their effect in electronic transport.  Our work supports that the large piezoresistive gauge in SmX materials stems from the sensitive deformation potential enabling thermal promotion of carriers from localized valence bands to delocalized conduction bands.  The piezoresistive gauge is driven by the change of the electronic band gap with strain as strain dependence of effective masses of valence and conduction bands is more than an order of magnitude smaller.  Previous bulk experimental data is reproduced by assuming Fermi level pinning near the valence band edge which gives optimal piezoresistive response.  Uniaxial strain can be employed to enhance the piezoresistive properties of Sm monochalcogenide compounds at the price of a larger component of stress along the compression axis (e.g. larger than the corresponding isotropic pressure).  For the uniaxial cases studied here, strain applied along the [001] direction exhibits the largest reduction in gap with volume change and thus a piezoresistive gauge larger than that of the isotropic case can be achieved.  We find that the sensitivity of SmX thin films can be altered by properly choosing the metal electrodes enabling optimization of the piezoresistive gauge by pinning the Fermi-level near the valence band.  Studies of  the complex band structure of SmSe show that the tunneling length can be smaller than 1 nm, indicating that the piezoresistive effect driven by thermal promotion of electrons (as observed in the bulk)  can be observed in films with thicknesses of a few nanometers. 

The authors acknowledge the support from DARPA's MESO Program under contract N66001-11-C-4109 and an IBM PhD fellowship award.  The use of nanoHUB.org computational resources operated by the Network for Computational Nanotechnology funded by the NSF under Grant Nos. EEC-0228390, EEC-1227110, EEC-0228390, EEC-0634750, OCI-0438246, and OCI-0832623 and OCI-0721680 is gratefully acknowledged. NEMO5 developments were critically supported by an NSF Peta-Apps award OCI-0749140 and by Intel Corp. The authors are grateful for fruitful discussion from Yaohua Tan.

\bibliography{report_piezo}

\begin{thebibliography}{47}
\bibitem [{uni()}]{uniaxial}%
  \BibitemOpen
  \href@noop {} {}\bibinfo {note} {If strain is not isotropic the top of the
  two valence bands split and problem cannot be solved
  analytically.}\BibitemShut {Stop}%
\bibitem [{not()}]{notesmte}%
  \BibitemOpen
  \href@noop {} {}\bibinfo {note} {The experimentally measured piezoresistive
  gauge in SmTe shows a kink at low pressures which the authors have been
  attributed to trap levels [9].}\BibitemShut {Stop}%
\bibitem [{ban()}]{bandmass}%
  \BibitemOpen
  \href@noop {} {}\bibinfo {note} {The band mass is computed employing the
  curvature of the electronic bands as obtained from
  first-principles.}\BibitemShut {Stop}%
\end{thebibliography}

\end{document}